# Observation of High-frequency Transverse Phonons in Metallic Glasses

X. Y. Li[1,2,3*], H. P. Zhang[4*], S. Lan[1,5*], D. L. Abernathy[6], T. Otomo[7], F. W. Wang[2,3,8], Y. Ren[9], M. Z. Li[4†], & X.-L. Wang[1, 10†]

[1]*Department of Physics, City University of Hong Kong, 83 Tat Chee Avenue, Hong Kong, China.*

[2]*Beijing National Laboratory for Condensed Matter Physics, Institute of Physics, Chinese Academy of Sciences, Beijing 100190, China.*

[3]*School of Physical Sciences, University of Chinese Academy of Sciences, Beijing 101408, China.*

[4]*Department of Physics, Beijing Key Laboratory of Opto-electronic Functional Materials and Micro-nano Devices, Renmin University of China, Beijing 100872, China.*

[5]*Herbert Gleiter Institute of Nanoscience, School of Materials Science and Engineering, Nanjing University of Science and Technology, 200 Xiaolingwei Avenue, Nanjing 210094, China.*

[6]*Neutron Scattering Division, Oak Ridge National Laboratory, Oak Ridge, Tennessee 37831, USA.*

[7]*Institute of Materials Structure Science, High Energy Accelerator Research Organization (KEK), Tsukuba, Ibaraki 305-0801, Japan.*

[8]*Songshan Lake Materials Laboratory, Dongguan 523808, China.*

[9]*X-ray Science Division, Argonne National Laboratory, Argonne, IL 60439, USA.*

[10]*Center for Neutron Scattering, City University of Hong Kong Shenzhen Research Institute, 8 Yuexing 1$^{st}$ Road, Shenzhen Hi-Tech Industrial Park, Shenzhen 518057, China.*

\* Co-first authors.

† Correspondence to: maozhili@ruc.edu.cn, xlwang@cityu.edu.hk

**Using inelastic neutron scattering and molecular dynamics simulations on a model Zr-Cu-Al metallic glass, we show that transverse phonons persist well into the high-frequency regime, and can be detected at large momentum transfer. Furthermore, the apparent peak width of the transverse phonons was found to follow the static structure factor. The one-to-one correspondence, which was demonstrated for both Zr-Cu-Al metallic glass and a 3-dimensional Lennard-Jones model glass, suggests a universal correlation between the phonon dynamics and the underlying disordered structure. This remarkable correlation, not found for longitudinal phonons, underscores the key role that transverse phonons hold for understanding the structure-dynamics relationship in disordered materials.**

*Introduction*. -Normal vibration modes and vibrational density of states (VDOS) are fundamental for understanding many of the physical properties of materials, such as dynamical excitations, and mechanical and thermal transport properties [1]. In crystalline materials, the normal modes, known as phonons, are quantized plane-wave solutions of the elemental modes of vibration [2], which can be well characterized experimentally by, *e.g.*, inelastic neutron scattering (INS) [3]. In amorphous materials, however, the phonon modes become far more complicated and are very different from those in crystals [4–13]. Most notably, the phonon spectra become broadened, and the extent of broadening depends on the degree of the disorder [14,15]. Still, in amorphous materials, phonon-like dispersions have been predicted, for example, by analytical theories for a 3-dimensional disordered system, first proposed by Hubbard and Beeby [4] and further developed by Takeno and Goda [6,16]. An extended review can be found in Yoshida and Takeno [17], with applications to both classical and quantum liquids. Calculations based on these theories show that the longitudinal mode has a strong dispersion, whereas the transverse mode is dispersionless at large wave vectors [4,6,16,17].

In amorphous materials, the transverse rather than the longitudinal phonon mode is more sensitive to disorder [7,18]. The transverse phonons are also responsible for many of the properties associated with the disorder. As an example, the velocity of transverse phonons is proportional to the square root of the shear modulus, which largely determines the deformation behavior of disordered materials [9,19]. Moreover, the transverse phonons have been linked with the Boson peak [7,8], or the excess VDOS, which is a universal feature in disordered materials. Recent simulations have found that for transverse phonons, the full-width-at-half-maxima (FWHM) of the phonon energy spectrum increases sharply with the momentum transfer $Q$ [7]. At a certain $Q$, the phonon wavelength and mean free path becomes comparable. This is the so-called Ioffe-Regel limit, beyond which the transverse phonons were believed to be extremely damped and therefore could no longer propagate [7].

Thus, it is of great interest, from both fundamental and practical perspectives, to understand transverse phonons, especially at large $Q$ where the effect of disorder becomes particularly pronounced. Yet, in spite of their importance, transverse phonons are difficult to determine experimentally [7,18,20]. For this reason, little is known about the nature of transverse phonons, *e.g.*, the contribution by different atomic species, or whether they exist at all. INS and lately inelastic x-ray scattering (IXS) are methods of choice for the experimental study of phonons [3,5,8,10,20–23]. While longitudinal phonons are straightforward to determine, measurements of transverse phonons require more considerations. In crystalline materials, the scattering intensity for transverse phonons vanishes in the 1$^{st}$ Brillouin zone, due to the scattering geometry factor, so the measurements of transverse phonons are typically done in the 2$^{nd}$ Brillouin zone and beyond [22,23]. In disordered materials, however, the disorder-induced phonon broadening [14] makes it difficult to determine transverse phonons due to the crossover of longitudinal and transverse branches. Thus, although a few IXS studies reported the measurements of transverse acoustic excitations in metallic liquids and glasses [5,24,25], the data are limited to small wave vectors (below $Q \sim 1$ Å$^{-1}$) and with significant uncertainties. INS is capable of reaching to larger $Q$, without suffering from the form factor drop-off as in IXS. Still, previous INS data have been plagued with inadequate precision due to the broad phonon spectrum and limited neutron beam flux [5,20].

The new chopper spectrometers at the state-of-the-art pulsed neutron sources provided opportunities to investigate the transverse phonons in disordered materials [26]. The high flux brought forth by the powerful sources, coupled with advances in neutron instrumentation, has enabled high-precision measurements to large $Q$, with fine energy resolution. Here, we report INS data for a ternary Zr-Cu-Al metallic glass (MG), where evidence of high-frequency transverse phonons was observed at an intermediate wave vector, around the boundary of the 2$^{nd}$ quasi Brillouin zone (QBZ) [20], see Fig. 1a for definition. The experimental observations are supported by molecular dynamics (MD) simulations, which not only confirms the existence of the transverse phonon branch at high-frequency but also shows how individual atoms contribute to the extended transverse mode. These findings present a challenge to the viewpoint that transverse phonons exist only at low-energies, below the Ioffe-Regel limit [7,18]. MD simulations answer the paradox, by showing that the transverse phonon peak width follows the static structure factor, rather than increasing monotonously with the phonon frequency.

*Experimental and simulation methods.* -Sample preparation and characterizations have been described elsewhere [27]. The INS experiment was carried out to measure the atomic dynamics in $Zr_{46}Cu_{46}Al_8$ MG, using the time-of-flight wide Angular-Range Chopper Spectrometer (ARCS) [28] at Spallation Neutron Source [29]. The measurements were performed with incident neutron energies of $E_i$ = 50 and 80 meV at room temperature (RT). For each $E_i$, the dynamic structure factor, $S(Q, E)$, where $Q$ and $E$ are the momentum and energy transfer, respectively, was generated using standard software MantidPlot [30]. 1-dimensional '$Q$-cuts' were taken along the $E$-axis to obtain the phonon spectra at specific $Q$-points by DAVE [31].

Classical MD simulations were also performed for $Zr_{46}Cu_{46}Al_8$ MG with the realistic embedded atom method (EAM) potential [32] using LAMMPS software package [33]. The glassy sample containing 10000 atoms at 300 K was obtained by fast quenching the equilibrated liquid at 2000 K with a cooling rate of $10^{12}$ K/s. Periodic boundary conditions were applied in 3 directions. In the cooling process, the isobaric and isothermal ensemble was employed and the sample size was adjusted to give a zero pressure. This was followed by the canonical ensemble MD at 300 K for data collection and analysis. The simulated static structure factor was benchmarked with experimental measurements [27]. A 3-dimensional Lennard-Jones binary glass was also simulated for comparison. The dynamic properties were analyzed in terms of van Hove correlation function (VHF), dynamic matrix method (DM), and velocity correlation function method (VCF), respectively. More details on simulation can be found in Supplemental Material [27].

*Results.* -Fig. 1a shows a representative INS data set of the $S(\boldsymbol{Q}, E)$ for $Zr_{46}Cu_{46}Al_8$ MG, obtained with $E_i = 50$ meV [27]. The full glassy nature of the samples was confirmed by the static structure factors, $S(Q)$, measured with neutron and synchrotron x-ray diffraction, shown in Fig. 1b.

The measurement results were analyzed in terms of the generalized $\boldsymbol{Q}$-dependent phonon density of states (GDOS), $G(\boldsymbol{Q}, E)$, which is related to $S(\boldsymbol{Q}, E)$, by the following equation [20,22,34],

$$G(\boldsymbol{Q}, E) = \left[1 - e^{-\frac{E}{k_B T}}\right] \frac{E}{Q^2} S(\boldsymbol{Q}, E) \qquad (1)$$

where $\left[1 - e^{-\frac{E}{k_B T}}\right]$ describes the Bose-Einstein statistics, $k_B$ is the Boltzmann constant and $T$ the temperature. The GDOS was reduced from the INS data using DAVE-mslice software [31]. Note that GDOS is slightly different from the current spectrum, defined as $C(\boldsymbol{Q}, E) = \frac{E^2}{Q^2} S(\boldsymbol{Q}, E)$, which has also been frequently used in analysis of the excitation spectrum in glasses and liquids [5]. The high-resolution experimental $\boldsymbol{Q}$-dependent GDOS (with $E_i$ = 50 meV) for $Zr_{46}Cu_{46}Al_8$ MG is presented in Fig. 2a. A characteristic dispersion relationship can be readily seen. The phonon spectrum is significantly broadened in energy, with the center band located at ~ 10 to 30 meV. The corresponding data for $Zr_{46}Cu_{46}Al_8$ MG obtained with $E_i$ = 80 meV are similar [27]. The phonon spectrum is dominated by the acoustic modes; no optical mode can be seen in Fig. 2a, even in $\boldsymbol{Q}$-dependent GDOS obtained with higher incident energies [27]. To examine the acoustic nature of the vibration modes, we performed Resonant Ultrasound Spectrometer (RUS) measurements [27]. The green and yellow solid lines in Fig. 2a are calculated dispersions at small $Q$ based on sound velocities measured by RUS, $V_L$ = 4218 m/s and $V_T$ = 2165 m/s, for longitudinal and transverse sound modes, respectively.

The dashed curves in Fig 2a are analytical calculations for dispersion relations in a disordered system [4,6,16,17,27], with the parameters of the Einstein frequency and inter-particle distance taken from experimental $G(\boldsymbol{Q}, E)$ and $S(Q)$, respectively [27]. Well separated longitudinal and transverse phonon branches are predicted by this model at QBZ boundaries and centers, *e.g.*, at $Q$ ~ 2.7 Å$^{-1}$ (the center) and $Q$ ~ 3.8 Å$^{-1}$ (the boundary) of the 2$^{nd}$ QBZ. Unfortunately, it is difficult to identify different branches from the experimental $\boldsymbol{Q}$-dependent GDOS, because the phonon spectra are too broad and the branches overlap. The simulated results analyzed by the VHF are shown in Fig. 2c. Here, a hint of branch separation can be seen at $Q$ ~ 3.8 Å$^{-1}$, the 2$^{nd}$ QBZ boundary. The blue stars superimposed in Fig. 2c are the longitudinal and transverse phonon peak positions extracted by fitting the $Q$-cut experimental GDOS at $Q \sim 3.8$ Å$^{-1}$ [27].

To confirm the separation of the longitudinal and transverse modes at $Q \sim 3.8$ Å$^{-1}$, the 2$^{nd}$-derivative method, a powerful data analysis technique to separate overlapping branches, was employed. This method, also known as the curvature method, has been successfully used in angle-resolved photoemission spectroscopy (ARPES) experiments to identify subtle features in the electronic density of states [35,36] from, *e.g.*, Weyl fermions in semimetals, and recently in electron energy loss spectroscopy experiments to identify the plasmons in a topological insulator [37]. In this study, the 2$^{nd}$-derivative with respect to $E$ was calculated and examined. The results are visualized in Figs. 2b, d in the vicinity of the 2$^{nd}$ QBZ, for the INS and MD simulated spectra. Two branches can be seen to separate starting from $Q \sim 3.5$ Å$^{-1}$. Linking together the experimental results, MD simulation, and the calculations by the analytical model, we can identify that at $Q \sim 3.8$ Å$^{-1}$, the upper branch at $E \sim 24$ meV corresponds to the longitudinal mode, whereas the lower branch at $E \sim 17$ meV corresponds to the transverse phonons.

In spite of the apparent success and expanding applications, caution must be exercised when applying the 2$^{nd}$-derivative method. A shift in peak position was sometimes noted when the 2$^{nd}$-derivative method was applied to the ARPES data. This was discussed in detail by Zhang *et al.* [38], and a correction method has been proposed. The peak produced by the 2$^{nd}$-derivative method may also be subject to large error, if the original data are of poor quality, *e.g.*, due to inadequate counting statistics. Naturally, the 2$^{nd}$-derivative method is most effective when discussed with reliable models, where the simulation results provide guidance on what the original and 2$^{nd}$-derivative data might look like.

To further verify and characterize the identified phonon modes, the DM and VCF methods were employed independently to compute the contributions of longitudinal and transverse branches. A comparison of the pros and cons of these three methods is discussed in Supplemental Material [27]. Overall, the calculations by different methods are consistent. The $Q$-dependent VDOS results calculated by the DM method are presented in Figs 2e, f, while those by the VCF method are shown in Supplemental Material [27]. Both methods show a dispersive longitudinal branch, while the transverse branch is flat beyond $Q \sim 2$ Å$^{-1}$, consistent with the calculations by analytical theories [4,6,16,17,27]. Distinctively, however, the transverse spectrum calculated by both methods exhibit a high concentration of strong intensities at $E \sim 17$ meV around $Q \sim 3.8$ Å$^{-1}$, exactly where the 2$^{nd}$-derivative of the experimental $Q$-dependent GDOS shows a separation.

More importantly, MD simulations by DM provided atomistic insights on the nature of the different phonon modes. In Fig. 3a, the VDOS and the contributions from individual atoms are plotted. We emphasize here that the VDOS and GDOS obtained in MD simulations are almost the same [27], because Cu and Zr have similar coherent scattering lengths and the contribution from Al is small. Thus, a comparison can be made between the calculated VDOS and experimental GDOS. Three peaks are identified in the MD simulations, which can be mapped to the experimental GDOS data. The MD simulations indicate that both Zr and Cu atoms contribute to the peaks at 17 and 24 meV. There is no contribution from Al atoms, however. This result also means that the experimentally observed phonon modes, at 17 and 24 meV (both transverse and longitudinal), involve Zr and Cu atoms only. On the other hand, the higher energy modes of 32 to 50 meV are mostly due to Al atoms.

To quantify the extent of a vibrational mode, the participation ratio $p(\omega_\lambda)$ was calculated as follows:

$$p(\omega_\lambda) = \frac{\left(\sum_j |e_\lambda(r_j)|^2\right)^2}{N \sum_j |e_\lambda(r_j)|^4} \tag{2}$$

where $e_\lambda(r_j)$ represents the normalized eigenvector of each atom $j$ in the mode $\omega_\lambda$. $p \approx 1/N$ corresponds to the localized modes, while $p \approx 1$ corresponds to the fully extended modes. Fig. 3b shows the $p$ of all vibrational modes. A $p$ value of ~50% for modes between 10 and 25 meV demonstrates that these vibrational modes are spatially extended [39]. They penetrate through the entire material, close to the nature of acoustic waves in solids. For the modes of 32 to 50 meV, which are solely due to Al atoms, the $p$ values ($<10^{-3}$) are very low and system size-dependent, indicating that these modes are indeed localized [9,40,41]. Thus, the mode of $E \sim 17$ meV is spatially extended, which is a necessary condition for propagating modes. To further elucidate the plane-wave feature of this mode at $E \sim 17$ meV and $Q \sim 3.8$ Å$^{-1}$, we analyzed the projection of the eigenstates on the plane waves over the whole $Q$ range. As shown in Fig. 3c, there exhibits strong $Q$-dependence in both longitudinal and transverse VDOS $D_\alpha(Q, E)$ ($\alpha \in L, T$) at $E = 17$ meV, indicating strong spatial correlation of this mode. The peaks of the $Q$-dependent $D_T(Q, E)$ indicate that there exist some characteristic $Q$ values in the transverse part of this mode, and one is located at $Q \sim 3.8$ Å$^{-1}$. This demonstrates the plane-wave character of the transverse phonon at $Q \sim 3.8$ Å$^{-1}$. In other words, the traditional transverse phonons still have its trace in high-frequency modes in disordered materials. We also compared $D_T(Q, E)$ for the modes of $E = 10$ and 23 meV whose $p$ values are also about 0.5. As shown in Fig. 3d, in contrast to the mode of $E = 17$ meV, the modes of $E = 10$ and 23 meV show much weaker $Q$-dependence.

This contrast reinforces our conclusion that the observed high-frequency transverse phonon at $E$ = 17 meV is indeed a spatially extended mode.

*Discussions*. -As disorder-induced broadening is a signature of the phonon spectrum in disordered materials [7,14,15], we have further studied the phonon width, or the lifetime, using MD simulations. The dispersion curves and peak widths were estimated by fitting the VDOS calculated by DM (Figs 2e, f) with a Gaussian function, following Arai *et al.* [20]. Fitting with a Lorentzian function gave similar results. Details of fitting are given in Supplemental Material [27]. We have found that for transverse phonons, the phonon width, measured by the FWHM, follows the $S(Q)$, as illustrated in Fig. 4a. Like the $S(Q)$, the FWHM reaches a minimum at the boundary of the $2^{nd}$ QBZ. This is where the spectrum for the transverse phonons is the sharpest, and hence the reason that the transverse phonons were observed at this $Q$ value rather than elsewhere. That being said, our study does not contradict the findings from ref. [7], which is limited to small $Q$, where the phonon peak width increases sharply with $Q$. Instead, our study points out that transverse phonons survive at larger $Q$. For the longitudinal mode, the phonon width also varies with $Q$, but there seems to be no correlation with $S(Q)$.

The correlation between the FWHM of transverse phonons and $S(Q)$ suggests that in amorphous materials, the phonon dynamics and static structure are intimately related. To check on whether this observation is specific to MGs, we have carried out a simulation with a generic Lennard-Jones potential, and the results are shown in Fig. 4b and Supplemental Material [27]. The remarkable one-to-one correspondence between the phonon FWHM and $S(Q)$ demonstrated here strongly suggests that the correlation is universal for amorphous materials. As in $Zr_{46}Cu_{46}Al_8$, the best point to observe the transverse phonons is also at the boundary of the 2$^{nd}$ QBZ, or the valley of $S(Q)$ following the first sharp diffraction peak, where the transverse phonons reveal themselves with the sharpest spectrum.

*Conclusion.* -Through high-quality INS measurements, we have established the existence of transverse acoustic phonons in amorphous materials. The transverse phonon modes, although significantly damped or broadened, persist well into the high-frequency regime, and beyond the 2$^{nd}$ QBZ. MD simulation provided the atomistic insights, showing that the transverse phonons are spatially extended, rather than localized, and involve only Zr and Cu atoms which form the short-range atomic clusters. Moreover, the simulations showed a universal one-to-one correspondence between the transverse phonon's width and the static structure factor, suggesting that in amorphous materials, the structure and dynamics are linked. It should be mentioned that the one-to-one correspondence seems to be unique to glass, as it is not found in crystalline materials. As transverse phonons are related to many fundamental phenomena in glass, such as the Boson peak and deformation behaviors, the findings of the present study should have broad implications in future study of the nature of glass and glass transitions.

**Acknowledgements:**

This work was supported by the Croucher Foundation (CityU Project No. 9500034) and Research Grant Council of Hong Kong SAR (JLFS/P-102/18). X.L.W. acknowledges support by the National Natural Science Foundation of China (NSFC) (No. 51571170) and the National Key R&D Program by the Ministry of Science and Technology (MOST) of China (No. 2016YFA0401501). S.L. acknowledges support by the National NSFC (No. 51871120) and the National Natural Science Foundation of Jiangsu Province (No. BK20171425). M.Z.L. acknowledges support by the National NSFC (No. 51631003) and MOST 973 Program (No. 2015CB856800). F.W.W. acknowledges support by the National NSFC (No. 11675255) and the National Key R&D Program by MOST of China (No. 2016YFA0401503). The neutron scattering experiments were carried out at the Spallation Neutron Source, which is sponsored by the Scientific User Facilities Division, Office of Basic Energy Sciences, US Department of Energy, under contract No. DE-AC05-00OR22725 with Oak Ridge National Laboratory.

**FIGURES**

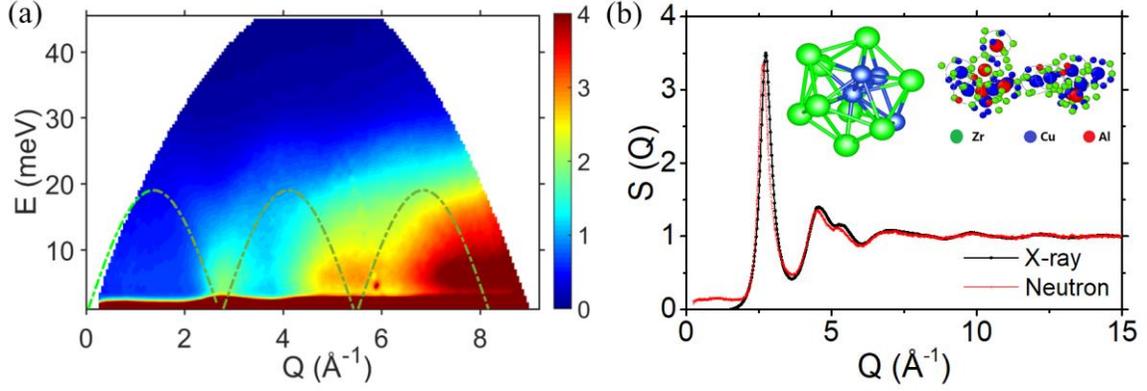

FIG. 1. The dynamic and static structure factors in $Zr_{46}Cu_{46}Al_8$ MG. (a) Dynamic structure factor, $S(Q, E)$, measured by INS with $E_i = 50$ meV on ARCS at RT. The dashed curve is calculated by the function, $E_{0*}\left|\sin\left(\pi\frac{Q}{Q_{max}}\right)\right|$, which defines the quasi Brillouin zone (QBZ), a term analogous to the Brillouin zone in crystalline materials in order to facilitate discussions of the phonon dynamics over different $Q$ regions. Here $Q_{max} = 2.74$ Å$^{-1}$ is the position of the first sharp diffraction peak and $E_0 = 19.02$ meV is an estimated Einstein frequency for vibration. (b) Static structure factor, $S(Q)$, measured by synchrotron x-ray and neutron total scattering at RT, demonstrated the fully amorphous nature of the sample. The inset shows a schematic configuration of the cluster-centered structure in $Zr_{46}Cu_{46}Al_8$ MG.

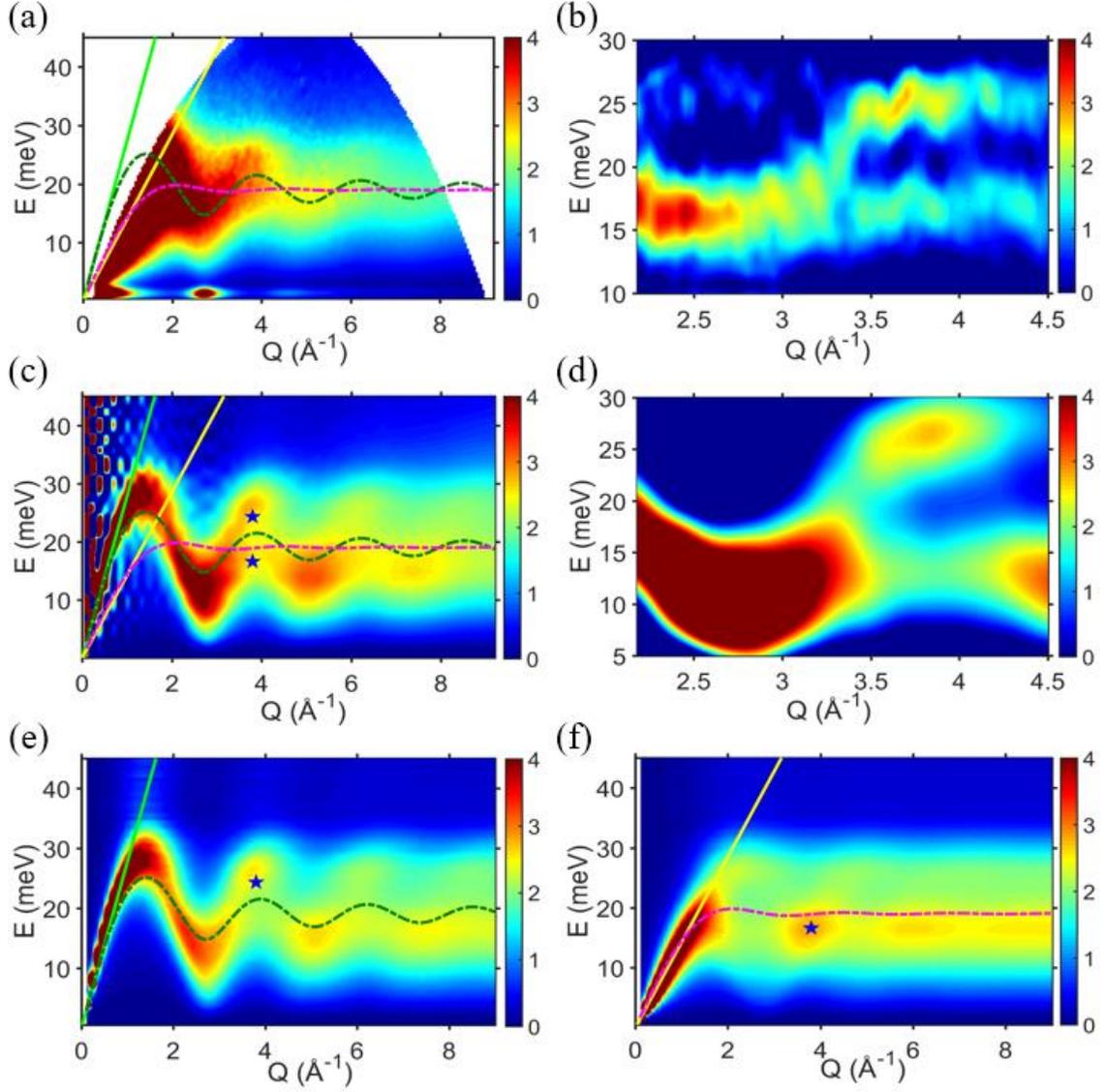

FIG. 2. Phonon-like dispersion relationship in $Zr_{46}Cu_{46}Al_8$ MG. (a and c) Phonon $Q$-dependent GDOS reduced from INS data and the corresponding neutron-weighted MD simulation results based on the VHF method. (b and d) Maps of the $2^{nd}$-derivatives of (a) and (c) in the vicinity of the $2^{nd}$ QBZ, respectively. Only the absolute values are plotted to trace the peaks, following the practice in refs. [35,36]. (e and f) The longitudinal and transverse $Q$-dependent VDOS obtained from MD simulations by the DM method. The results clearly show that the transverse branch has an intensity maximum around $Q = 3.8$

Å$^{-1}$. The olive and magenta curves in (a, c, e and f) are calculated longitudinal and transverse phonon dispersions based on the analytical theory for a disordered system [4,6,16,17,27]. The green and yellow lines in (a, c, e and f) are calculated dispersions based on longitudinal and transverse sound velocities measured by the RUS method. These data are consistent with each other in the low $Q$ regime. The blue stars in (c, e and f) marks the boundary of the 2$^{nd}$ QBZ, where the transverse phonons manifest. For the purpose of comparison, the color bars are plotted in relative intensities.

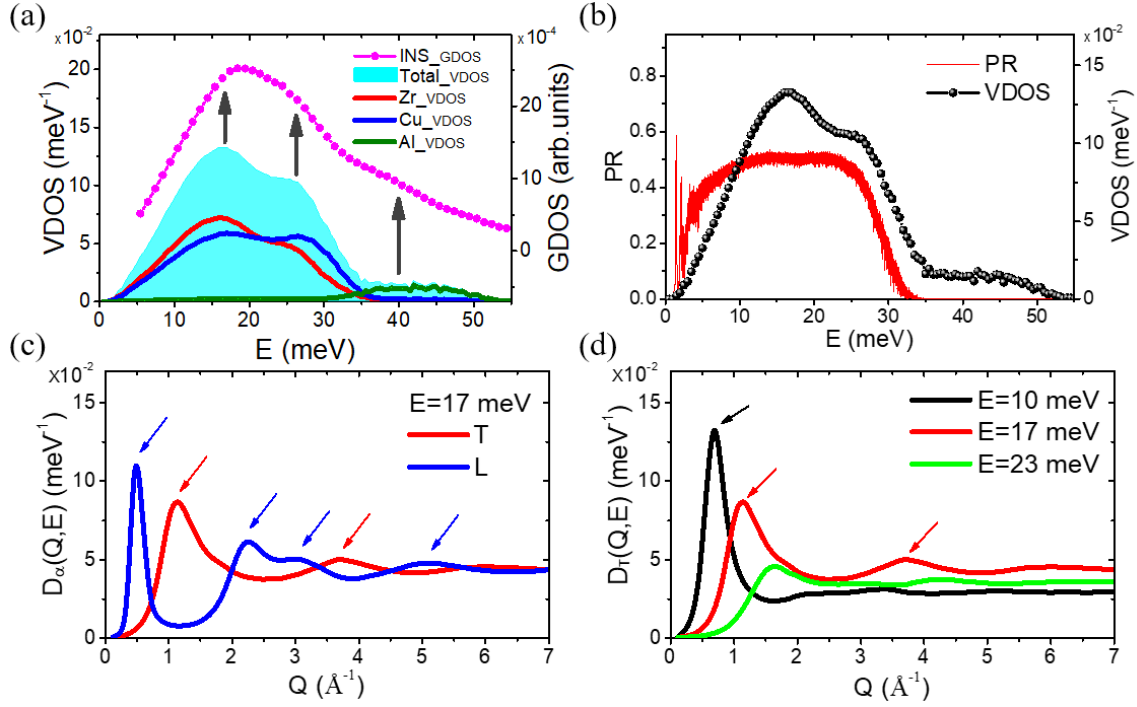

FIG. 3. MD simulation results based on the DM method. (a) The total VDOS, and partial VDOS for each element, and (b) the participation ratio (PR) for each mode. For comparison, the experimental GDOS is superimposed in (a), which was obtained by averaging a $Q$ range of 2.5 – 9.0 Å$^{-1}$ for the INS data. (c) $Q$-dependence of the transverse and longitudinal VDOS $D_\alpha(Q,E)$ ($\alpha \in L, T$) at $E$ = 17 meV. These peaks indicate the plane-wave characteristics in this mode. (d) Comparison of $D_T(Q,E)$ at $E$ = 10, 17, and 23 meV. The mode of $E$ = 17 meV shows stronger transverse wave characters around $Q$ = 3.8 Å$^{-1}$.

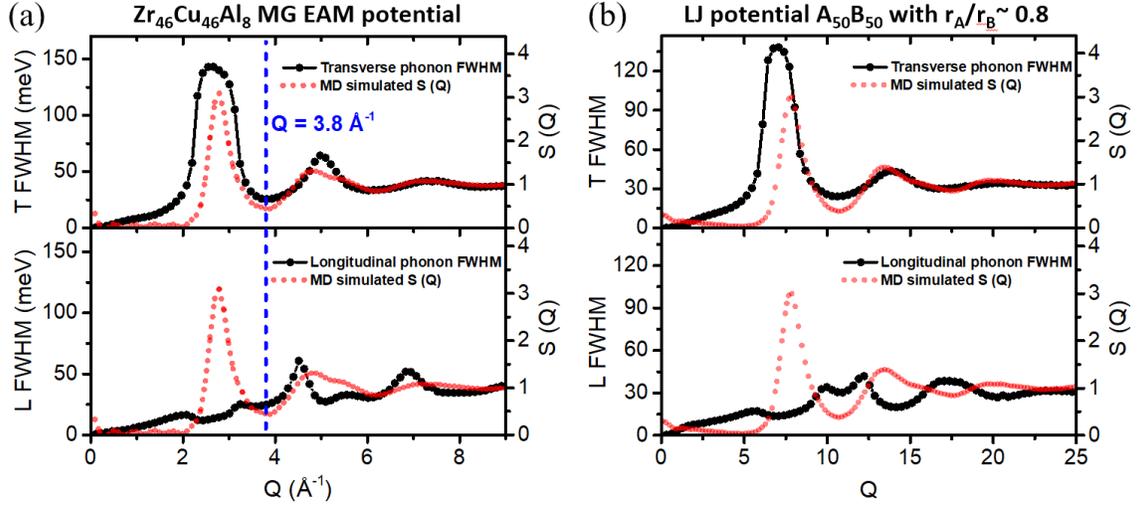

FIG. 4. Peak widths of the longitudinal and transverse phonon spectra vs. $S(Q)$ obtained by MD simulations based on the DM method. The widths were obtained by fitting the MD spectra with a Gaussian function [20,27]. (a) Empirical potential for $Zr_{46}Cu_{46}Al_8$ MG. (b) Lennard-Jones (LJ) potential for a model glass. A one-to-one correlation can be seen between the apparent peak width and $S(Q)$ for the transverse phonon mode. These results suggest a universal correlation between the transverse phonon dynamics and the underlying disordered structure. No correlation was found for the longitudinal phonons. These LJ results are in scaled units.

# SUPPLEMENTAL MATERIAL:

# Observation of High-frequency Transverse Phonons in Metallic Glasses

X. Y. Li[1,2,3*], H. P. Zhang[4*], S. Lan[1,5*], D. L. Abernathy[6], T. Otomo[7], F. W. Wang[2,3,8], Y. Ren[9], M. Z. Li[4†], & X.-L. Wang[1, 10†]

\* Co-first authors.

† Correspondence to: maozhili@ruc.edu.cn, xlwang@cityu.edu.hk

**CONTENTS**

## S1. Sample Preparation and Characterization

**Sample Synthesis.** Amorphous alloy ingots with compositions of $Zr_{46}Cu_{46}Al_8$ were prepared by arc melting a mixture of Zr (99.99%), Cu (99.99%), and Al (99.99%) in appropriate amounts under a Ti-gettered argon atmosphere. Each ingot was re-melted six times to ensure compositional homogeneity and quenched into cooper-mold by suction casting under high purity Ar atmosphere.

**Resonant Ultrasound Spectrometer (RUS) Measurements.** Elastic constants of the $Zr_{46}Cu_{46}Al_8$ MG, including Young's modulus, shear modulus, and sound velocity for longitudinal $V_L$ and transverse $V_T$, were measured using RUS at room temperature (RT). Rectangle samples about 2 mm × 2 mm × 4 mm with known volume and mass were placed between the piezoelectric transducers, and the two independent elastic constants $C_{11}$ and $C_{44}$ for each alloy were obtained and used to calculate the elastic moduli and sound velocity.

**Synchrotron x-ray and Neutron Diffraction Measurements.** High-energy synchrotron x-ray total scattering measurements were carried out at the beamline 11-ID-C at the Advanced Photon Source, Argonne National Laboratory in USA. High-energy x-rays with a wavelength of 0.10804 Å and a beam size of 0.5 mm × 0.5 mm were used in transmission geometry for data collection at RT. Neutron total scattering measurements were carried out at RT at the beamline BL21 NOVA of J-PARC in Japan.

## S2. Molecular Dynamics (MD) Simulations

In this work, classical MD simulations were performed to investigate the structural and dynamic characteristics of ternary $Zr_{46}Cu_{46}Al_8$ MG within the LAMMPS package [1,2]. In the $Zr_{46}Cu_{46}Al_8$ system, the interatomic interactions were described by the realistic embedded-atom method (EAM) potential developed by Sheng *et al.* [3]. In brief, the potential energy of atom *i* is divided into two contributions: a pairwise part and a local density part:

$$E_i = F_\alpha\left(\sum_{j\neq i} \rho_\beta(r_{ij})\right) + \frac{1}{2}\sum_{j\neq i} \phi_{\alpha\beta}(r_{ij})$$

where *α* and *β* are the element types of the center (embedded) atom *i* and the neighboring atom *j*. $\rho_\beta$ is the charge density function of *β*, $F_\alpha$ is the embedding function of *α*, $\phi_{\alpha\beta}$ is the pair potential function between *α* and *β*. The total potential energy of the system is the summation of the potential energy of each atom. We simulated the system containing 10000 atoms in a cubic box with periodic boundary conditions applied in three directions. In the process of sample preparation, it was first melted and equilibrated at 2000 K for 2.0 ns (MD time step is 2.0 fs) followed by hyper-quenching to 300 K with a cooling rate of $10^{12}$ K/s, then relaxed for 2.0 ns at 300 K to reach its equilibrium. In this process, the isobaric and isothermal (NPT) ensemble was used with the sample size being adjusted to give zero pressure. After that, the canonical (NVT) ensemble MD was conducted at 300 K for data collections and analysis. As shown in FIG. SI-1, the simulated static structure factor of $Zr_{46}Cu_{46}Al_8$ MG was benchmarked with experimental measurements. A good agreement is demonstrated.

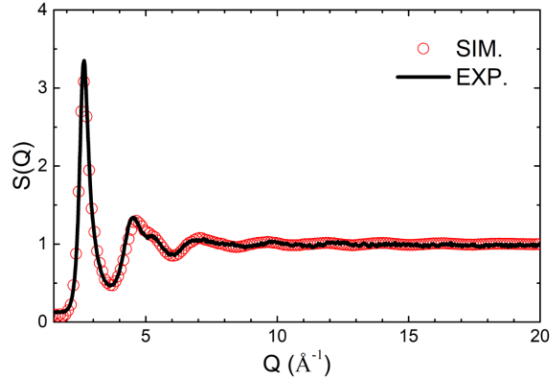

FIG. SI-1. Comparison of the static structure factor of $Zr_{46}Cu_{46}Al_8$ MG obtained from the neutron experiments (EXP) and MD simulations (SIM).

We also performed MD simulations for 3-dimensional Lennard-Jones binary mixtures composed of two atomic species, A and B, with $N_A = N_B = 5000$ particles. The interaction potential has the Lennard-Jones functional form:

$$U_{ij} = 4\varepsilon\left[\left(\frac{\sigma_{ij}}{r_{ij}}\right)^{12} - \left(\frac{\sigma_{ij}}{r_{ij}}\right)^{6}\right]$$

where $\sigma_{ij} = (\sigma_i + \sigma_j)/2$ and $\sigma_i$ is the diameter of particle $i$ ($i, j$ = A, B). We used particle mass $m_B$, $\sigma_B$, $\varepsilon$, and $\tau_0 = \sqrt{m_B \sigma_B^2/\varepsilon}$ as the basic unit of mass, length, energy and time, respectively. The mass ratio is $m_A/m_B = 1.0$ and the size ratio is $\sigma_A/\sigma_B = 0.8$. The liquid at $T = 0.6$ was equilibrated for $1000\tau_0$ and then quenched to $T = 0.1$ at a cooling rate of $1\times10^{-4}$ under zero pressure. At $T = 0.1$, the sample was further annealed for $1000\tau_0$ to reach its equilibrium. After that, the NVT ensemble MD was conducted for data collections.

**S3. $S(Q, E)$ and $G(Q, E)$ Measured by INS with $E_i$ = 80 meV**

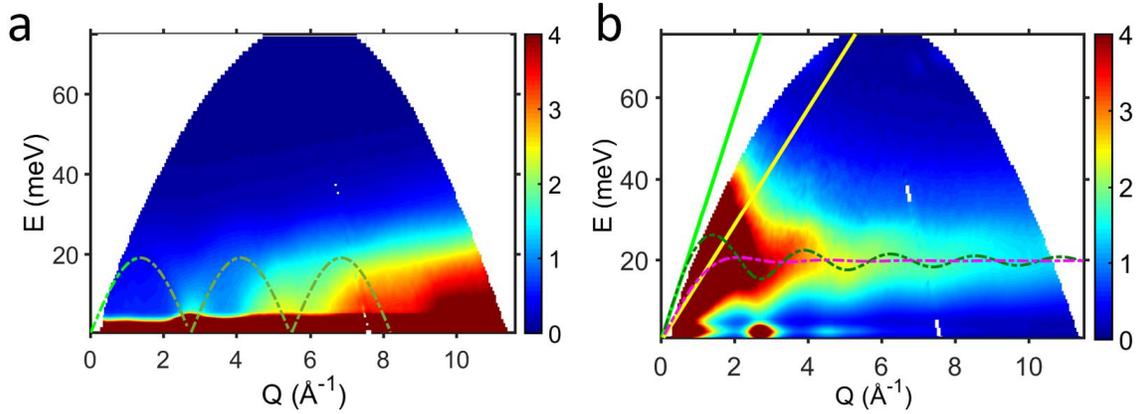

FIG. SI-2. Dynamic structure factor $S(Q, E)$ (a) and general $Q$-dependent density of states $G(Q, E)$ (b) in $Zr_{46}Cu_{46}Al_8$ MG measured by INS at RT with $E_i$ = 80 meV. The dashed curve in (a) was obtained by the function of $E_{0*}\left|\sin\left(\pi\frac{Q}{Q_{max}}\right)\right|$, which defines the quasi Brillouin zone (QBZ), analogous to the Brillouin zone in crystalline materials. In (b), the

olive and magenta curves are longitudinal and transverse phonon dispersion relationships calculated by the analytical theory for a disordered system [see S4]. The green and yellow solid lines are calculated dispersions based on longitudinal and transverse sound velocities measured by the RUS method, which are in good agreement with the analytical theory.

## S4. Analytical Theory of Phonon Dispersion in Disordered Materials

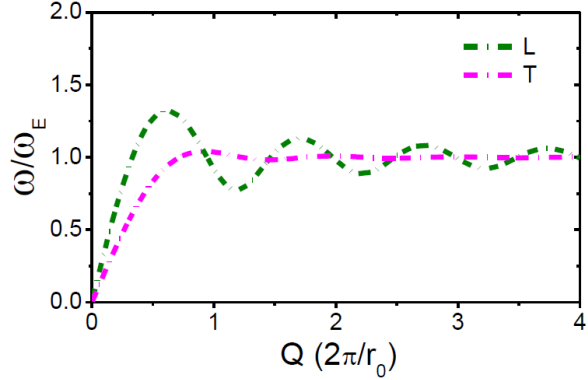

FIG. SI-3. The longitudinal and transverse dispersion relationships predicted by Eq.(4) & Eq.(5), respectively. For $Zr_{46}Cu_{46}Al_8$ MG, $\omega_E$ and $r_0$ were estimated based on the experimental $G(Q, E)$ and $S(Q)$, respectively (see S5 and S6).

Analytical theories for phonons in three-dimensional disordered systems were first proposed by Hubbard and Beeby [4] and further developed by Takeno and Goda [5,6]. An extended review can be found in Yoshida and Takeno [7], with applications to both classical and quantum liquids. The vibration modes in disordered materials were assumed to be phonon-like plane waves, so that the dispersion relation can be derived by diagonalization of the dynamic matrix in reciprocal space $\mathcal{D}_{\alpha\beta}(\vec{k})$ as:

$$\det|\omega^2\delta(\alpha\beta) - \mathcal{D}_{\alpha\beta}(\vec{k})| = 0 \quad (1)$$

where $\omega$ is the eigenfrequency of the phonon-like excitations. Assuming that the interactions could be described as an effective pair potential, the dynamic matrix in reciprocal space of an isotropic amorphous system can be written in the integral form,

$$\mathcal{D}_{\alpha\beta}(\vec{k}) = \frac{\rho}{M}\int d\vec{r}\, g(r)\nabla_\alpha\nabla_\beta\langle V(r)\rangle[1 - \exp\{i\vec{k}\cdot\vec{r}\}] \quad (2)$$

where $g(r)$ is the pair correlation function, and $\nabla_\alpha\nabla_\beta\langle V(r)\rangle$ (known as the force constant) is the second order partial derivative of the pair potential. $\rho$ and $M$ are the atomic number density and atomic mass, respectively. By taking a simple approximation of:

$$g(r)\nabla_\alpha\nabla_\beta\langle V(r)\rangle = \delta(r - r_0) \quad (3)$$

where $r_0$ is the nearest-neighbor atomic distance, Eq.(2) can be solved analytically. The dispersion relationships can be derived as:

$$\omega_L^2(Q) = \omega_E^2 \left[1 - \frac{3\sin(Qr_0)}{Qr_0} - \frac{6\cos(Qr_0)}{(Qr_0)^2} + \frac{6\sin(Qr_0)}{(Qr_0)^3}\right] \quad (4)$$

$$\omega_T^2(Q) = \omega_E^2 \left[1 + \frac{3\cos(Qr_0)}{(Qr_0)^2} - \frac{3\sin(Qr_0)}{(Qr_0)^3}\right] \quad (5)$$

where $\omega_E$ is termed as the Einstein frequency. FIG. SI-3 shows the longitudinal and transverse phonon-like dispersion relationships predicted by Eq.(4) & Eq.(5), respectively. For $Zr_{46}Cu_{46}Al_8$ MG, $\omega_E$ and $r_0$ were estimated based on the experimental $G(Q, E)$ and $S(Q)$, respectively (See S5 and S6 for details). The longitudinal phonons show a strong dispersion relationship, whereas the transverse phonons are flat beyond $Q \sim 1$.

## S5. Determination of the Nearest-neighbor Atomic Distance

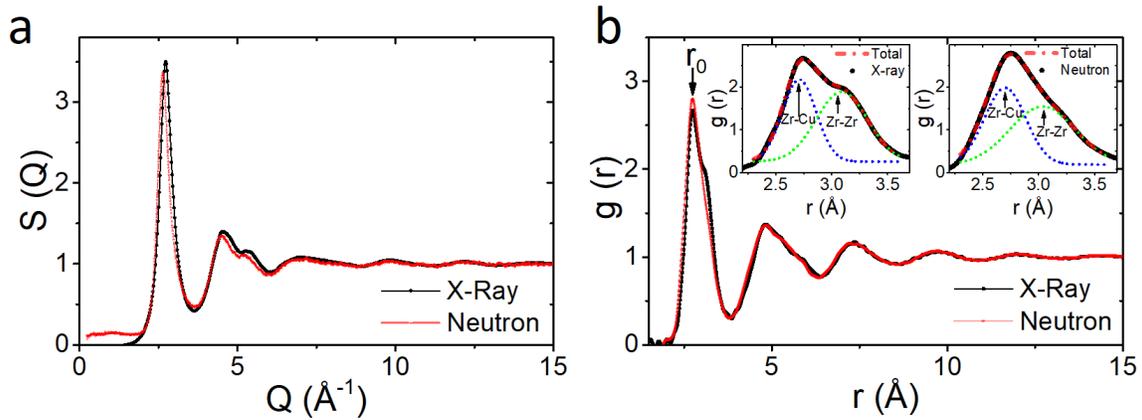

FIG. SI-4. The static structure factor $S(Q)$ (a) and the corresponding pair distribution function (PDF) $g(r)$ (b) in $Zr_{46}Cu_{46}Al_8$ MG measured at RT by synchrotron x-ray and neutron diffraction, respectively. The PDF data were analyzed by PDFgetX [8] and PDFgetN [9]. The extracted nearest-neighbor atomic distance is $r_0 = 2.74$ Å.

## S6. Determination of Einstein Frequency from GDOS

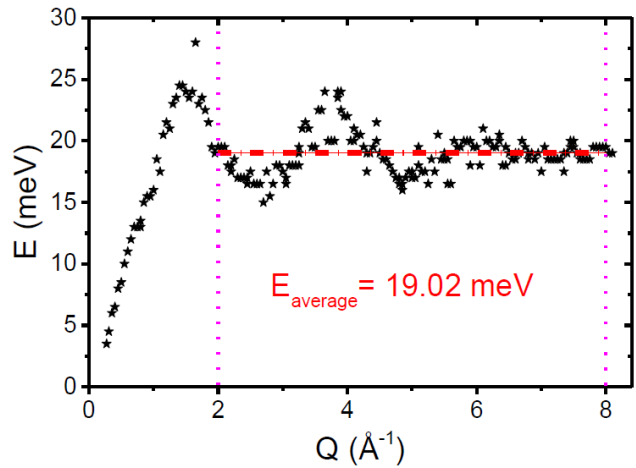

FIG. SI-5. The peak positions of $Q$-dependent GDOS at each fixed $Q$ in $Zr_{46}Cu_{46}Al_8$ MG measured by INS with $E_i = 50$ meV at RT. The average from 2 to 8 Å$^{-1}$ yielded a value of 19.02 meV for the Einstein Frequency $\omega_E$.

In the analytical theory equation for dispersion relationships, the phonon frequency reaches an asymptotic value, or the Einstein frequency, as $Q$ goes to infinity. In practice, however, due to the limited $Q$ range in experimental measurements, the Einstein frequency was estimated by averaging the frequency at large $Q$ values, from 2 to 8 Å$^{-1}$, as shown in FIG.SI-5. The value was found to be $E_0 = 19.02$ meV.

## S7. Determination of Longitudinal and Transverse Phonon Peak Positions

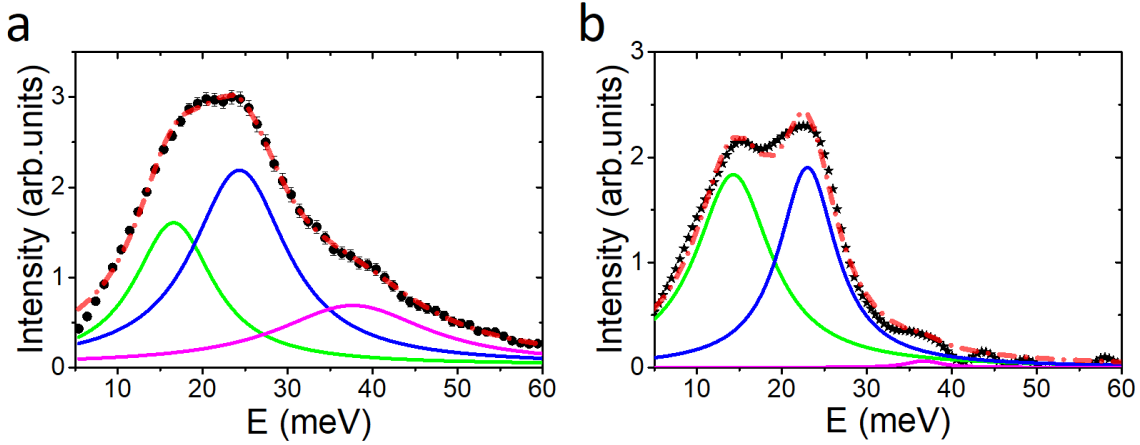

FIG. SI-6. GDOS at $Q \sim 3.8$ Å$^{-1}$ obtained from INS experiments (a) and MD simulations (b), respectively (dotted curves). Solid curves represent individual components from (a) Voigt function fitting and (b) Lorentzian function fitting, respectively. Dash-dotted lines in (a) and (b) are the overall fitting results, which are the sum of three solid curves. The fitting parameters are listed in Table SI-1.

The Voigt function $f_V$, a convolution of a Lorentzian and a Gaussian function, which is often adopted to fit the experimental GDOS, is defined as:

$$f_V = C_0 + (f_L * f_G)(x) = C_0 + A \frac{2\ln2}{\pi^{3/2}} \frac{\sigma_L}{\sigma_G^2} \int_{-\infty}^{\infty} \frac{e^{-t^2}}{\left(\sqrt{\ln2}\frac{\sigma_L}{\sigma_G}\right)^2 + \left(\sqrt{4\ln2}\frac{x-x_c}{\sigma_G}-t\right)^2} dt \quad (6)$$

Here Lorentzian function $f_L$ and Gaussian function $f_G$ are defined as:

$$f_L = \frac{2A}{\pi} \frac{\sigma_L}{4(x-x_c)^2 + \sigma_L^2} \qquad f_G = \sqrt{\frac{4\ln2}{\pi\sigma_G^2}} exp\left(-\frac{4\ln2}{\sigma_G^2}x^2\right) \quad (7)$$

where $x_c$ is the peak center, $\sigma_L$ is the peak width, $A$ is amplitude parameter and $C_0$ is background, and $\sigma_G$ is the instrument resolution parameter.

The agreement factor $R$-square is given by,

$$R^2 = 1 - \frac{\sum_{i=1}^{n}(y_i - y_{cal})^2 / df_{Error}}{\sum_{i=1}^{n}(y_i - \bar{y})^2 / df_{Total}}$$

where $y_i$ is the experimental data, and $y_{cal}$ is the calculation data, and $\bar{y}$ is the experimental data average value, with degree of freedom of error ($df_{Error}$) and degree of freedom of total ($df_{Total}$).

Table. SI-1. Fitting parameters in determining longitudinal and transverse phonon peak positions at $Q \sim 3.8$ Å$^{-1}$.

|  | Parameters | Experiment | Simulation |
|---|---|---|---|
| Transverse | $x_c$ / meV | 16.6 (±0.5) | 14.3 (±0.2) |
|  | $\sigma_L$ / meV | 11 (±1) | 10.7 (±0.5) |
|  | $A$ | 27 (±7) | 31 (±2) |
| Longitudinal | $x_c$ / meV | 24.3 (±0.4) | 23.0 (±0.2) |
|  | $\sigma_L$ / meV | 13 (±2) | 8.2 (±0.5) |
|  | $A$ | 45 (±11) | 25 (±2) |
|  | $R^2$ | 0.997 | 0.987 |

Here GDOS at 3.8 Å$^{-1}$ measured by $E_i = 80$ meV was fitted with three Voigt functions, which correspond to transverse phonon, longitudinal phonon, and a high-frequency Al-dominated local mode (see FIG.3), respectively. The energy resolution for ARCS is around 2 meV at $E \sim 30$ meV, so we set $\sigma_G = 2$ meV in the fitting process. For comparison, the GDOS obtained by MD simulations was also fitted here with three Lorentzian functions (without considering the instrumental resolution). As shown in FIG. SI-6, the GDOS were well described by the fitting functions over the whole energy range (from 0 to 60 meV). The fitting parameters, $x_c$, $\sigma_L$ and $A$, are listed in table SI-1. The fitted (experimental data) transverse and longitudinal phonons energies were 16.6 meV and 24.3 meV, respectively, and marked with blue stars in FIG. 2 and FIG. SI-8.

## S8. Vibrational Density of States (VDOS) Analysis for Simulated Zr$_{46}$Cu$_{46}$Al$_8$ MG

To calculate $\boldsymbol{Q}$-dependent VDOS, three different methods were adopted: van Hove correlation function (VHF) [10,11], dynamic matrix (DM) [12,13], and velocity correlation function (VCF) [12,13]. The VHF and VCF were calculated based on the NVT runs of the glass sample at low temperature, whereas the DM was obtained from the inherent structure, which is obtained by minimizing the final glass sample.

**van Hove Correlation Function (VHF) Method.** The VHF, also known as the dynamic pair correlation function, is defined as the probability for particle $k$ at time $t = 0$ to find a particle $j$ at a distance $r$ after a certain time $t$,

$$G(\boldsymbol{r}, t) = \frac{1}{N} \langle \sum_j^N \sum_k^N \delta(\boldsymbol{r} + \boldsymbol{r}_{j,t} - \boldsymbol{r}_{k,0}) \rangle$$

where $\langle\ \rangle$ donates the ensemble average, $N$ is the total number of particles, $\boldsymbol{r}_{j,t}$ and $\boldsymbol{r}_{j,0}$ represent the position vectors of atom $j$ at time $t$ and 0, respectively.

The dynamic structure factor, which can be measured in INS experiments, can be calculated directly via Fourier transformation of VHF. The neutron-weighted coherent and incoherent VHFs can be expressed as:

$$b_{coh}^2 \cdot G_{coh}(\boldsymbol{r},t) = \frac{1}{N} \langle \sum_j^N \sum_k^N b_{j,coh} \cdot b_{k,coh} \cdot \delta(\boldsymbol{r} + \boldsymbol{r}_{j,t} - \boldsymbol{r}_{k,0}) \rangle$$

$$b_{inc}^2 \cdot G_{inc}(\boldsymbol{r},t) = \frac{1}{N} \langle \sum_j^N b_{j,inc}^2 \cdot \delta(\boldsymbol{r} + \boldsymbol{r}_{j,t} - \boldsymbol{r}_{j,0}) \rangle$$

Here $b_{j,inc}$ and $b_{j,coh}$ are the incoherent and coherent neutron scattering lengths of atom $j$, while $b_{inc}$ and $b_{coh}$ are the average incoherent and coherent neutron scattering lengths which can be calculated accordingly:

$$b_{coh}^2 = \left(N^{-1} \sum_j b_{j,coh}\right)^2 \qquad b_{inc}^2 = N^{-1} \sum_j b_{j,inc}^2$$

Thus, the coherent and incoherent neutron-weighted dynamic structure factors can be derived:

$$S_\alpha(\boldsymbol{Q}, \omega) = \frac{1}{2\pi} \iint G_\alpha(\boldsymbol{r},t) \exp(i\boldsymbol{Q} \cdot \boldsymbol{r} - i\omega t) d\boldsymbol{r} dt, (\alpha \in coh, inc)$$

The double differential scattering cross section $\frac{d^2\sigma}{d\Omega d\omega}$ measured in INS experiments is a superposition of coherent and incoherent parts.

$$\frac{k_i}{k_f} \frac{d^2\sigma}{d\Omega d\omega} = N[b_{coh}^2 \cdot S_{coh}(\boldsymbol{Q}, \omega) + b_{inc}^2 \cdot S_{inc}(\boldsymbol{Q}, \omega)]$$

where $k_i$ and $k_f$ are, respectively, magnitudes of the wave vectors corresponding to incident and scattered neutrons. In $Zr_{46}Cu_{46}Al_8$ system, the coherent scattering lengths of Zr, Cu, and Al elements are 7.16, 7.718, and 3.449 fm, respectively, and the incoherent scattering cross sections ($4\pi b_{inc}^2$) of three elements are 2, 55, and 0.82 fm$^2$ [14], respectively. Thus, the $\boldsymbol{Q}$-dependent GDOS can be calculated through equation (1) in the manuscript. The calculated neutron-weighted phonon $\boldsymbol{Q}$-dependent GDOS can be compared directly with that obtained from INS experiments, as shown in FIG. 2.

**Dynamic Matrix (DM) Method.** In the DM method, the real space DM of the inherent structure (configuration corresponding to the potential energy minimum) is given by:

$$\boldsymbol{D}_{j,k} = \frac{1}{\sqrt{M_j M_k}} \frac{\partial^2 U(x_1, y_1, z_1, \cdots, x_N, y_N, z_N)}{\partial R_j \partial R_k}$$

where $M_j$ is mass of atom $j$ and $R_j$ is the coordinate ($x$, $y$ or $z$) of atom $j$. We directly diagonalize the DM and calculate the $\boldsymbol{Q}$-dependent VDOS as:

$$D_{DM}(\mathbf{Q},\omega) = \sum_{\lambda} |\mathbf{e}_\lambda(\mathbf{Q})|^2 \delta(\omega - \omega_\lambda)$$

where $\omega_\lambda$ is the eigenfrequency, and $\mathbf{e}_\lambda(\mathbf{Q}) = \sum_j \mathbf{e}_\lambda(\mathbf{r}_j) exp(i\mathbf{Q} \cdot \mathbf{r}_j)$ is the projection of the eigenstate onto the plane waves. Considering the relative direction of $\mathbf{e}_\lambda(\mathbf{Q})$ to the wave vector ($\hat{\mathbf{Q}} = \mathbf{Q}/|\mathbf{Q}|$), the $\mathbf{Q}$-dependent VDOS can be separated into longitudinal and transverse parts as:

$$\mathbf{e}_{\lambda,L}(\mathbf{Q}) = [\mathbf{e}_\lambda(\mathbf{Q}) \cdot \hat{\mathbf{Q}}]\hat{\mathbf{Q}}$$
$$\mathbf{e}_{\lambda,T}(\mathbf{Q}) = \mathbf{e}_\lambda(\mathbf{Q}) - \mathbf{e}_{\lambda,L}(\mathbf{Q})$$

so that

$$D_{DM,\alpha}(\mathbf{Q},\omega) = \sum_{\lambda} |\mathbf{e}_{\lambda,\alpha}(\mathbf{Q})|^2 \delta(\omega - \omega_\lambda), (\alpha \in L, T)$$

**Velocity Correlation Function (VCF) Method.** In the VCF method, the $\mathbf{Q}$-dependent VDOS can be calculated by the Fourier transform of the VCF as:

$$D_{FT}(\mathbf{Q},\omega) = \frac{1}{2\pi} \int \langle \mathbf{j}(\mathbf{Q},t) \cdot \mathbf{j}(-\mathbf{Q},0) \rangle exp(-i\omega t) dt$$

in which $\mathbf{j}(\mathbf{Q},t) = \sum_j \frac{\mathbf{V}_{j,t}}{\sqrt{k_B T/M_j}} exp(i\mathbf{Q} \cdot \mathbf{r}_{j,t})$ ($\mathbf{V}_{j,t}$ is the velocity of atom $j$ at time $t$) is the projection of the velocity field onto the plane waves at time $t$. Similarly, the $\mathbf{Q}$-dependent VDOS make up of longitudinal and transverse parts as:

$$\mathbf{j}_L(\mathbf{Q},t) = [\mathbf{j}(\mathbf{Q},t) \cdot \hat{\mathbf{Q}}]\hat{\mathbf{Q}}$$
$$\mathbf{j}_T(\mathbf{Q},t) = \mathbf{j}(\mathbf{Q},t) - \mathbf{j}_L(\mathbf{Q},t)$$
$$D_{FT,\alpha}(\mathbf{Q},\omega) = \frac{1}{2\pi} \int \langle \mathbf{j}_\alpha(\mathbf{Q},t) \cdot \mathbf{j}_\alpha(-\mathbf{Q},0) \rangle exp(-i\omega t) dt, (\alpha \in L, T)$$

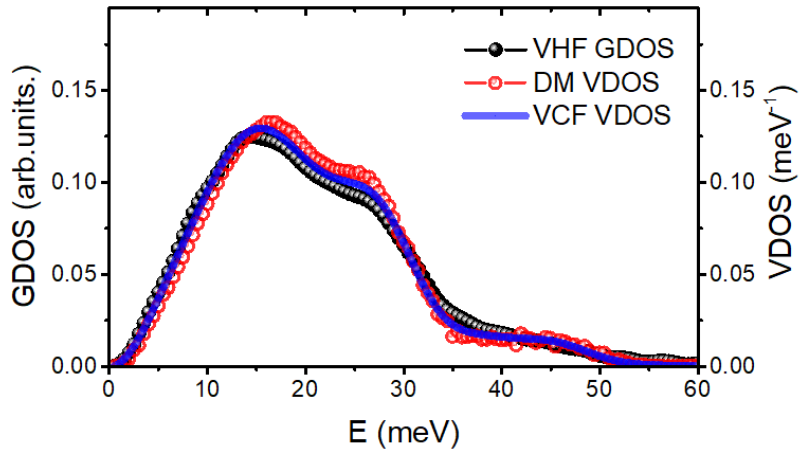

FIG. SI-7. A comparison of DOS of $Zr_{46}Cu_{46}Al_8$ MG calculated with VHF, DM and VCF methods, respectively.

As shown above, the VHF method is convenient to calculate the dynamic structure factors and make a direct comparison with experimental measurements. However, one cannot obtain specific information concerning the transverse or longitudinal phonons from the $Q$-dependent GDOS calculated by VHF. While partial $Q$-dependent VDOS of longitudinal and transverse phonons can be directly obtained by DM and VCF methods, experimental parameters such as scattering cross-section cannot be incorporated into the calculations. Thus, the $Q$-dependent VDOS was comprehensively analyzed by three methods in our numerical calculations. FIG. SI-7 shows the good agreement in DOS obtained by VHF, DM and VCF methods, respectively.

## S9. $Q$-dependent VDOS Calculated by VCF

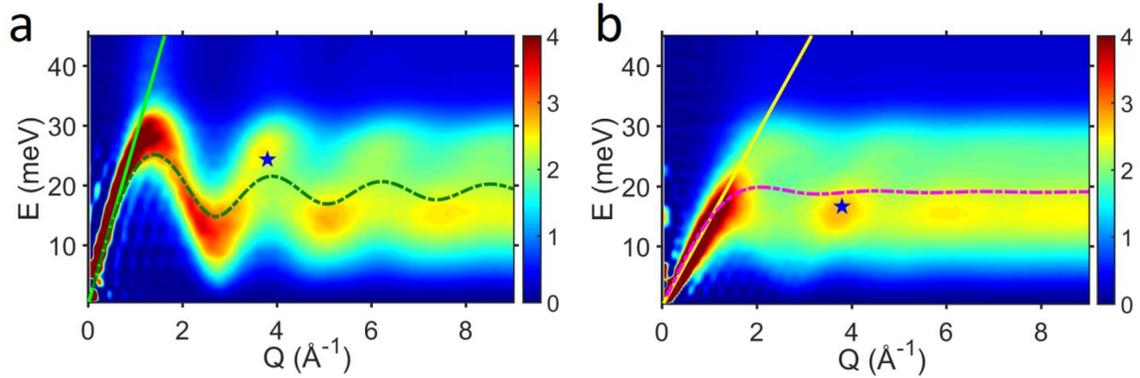

FIG. SI-8. $Q$-dependent VDOS of (a) longitudinal and (b) transverse phonons in $Zr_{46}Cu_{46}Al_8$ MG calculated by VCF. The curves and points are kept the same as in FIGs. 2(e) and 2(f), to allow for easy comparison. The calculated partial $Q$-dependent VDOS by the VCF method here is consistent with the DM results shown in FIGs. 2(e) and 2(f).

## S10. Determination of FWHM of Longitudinal and Transverse Phonons in $Zr_{46}Cu_{46}Al_8$ MG

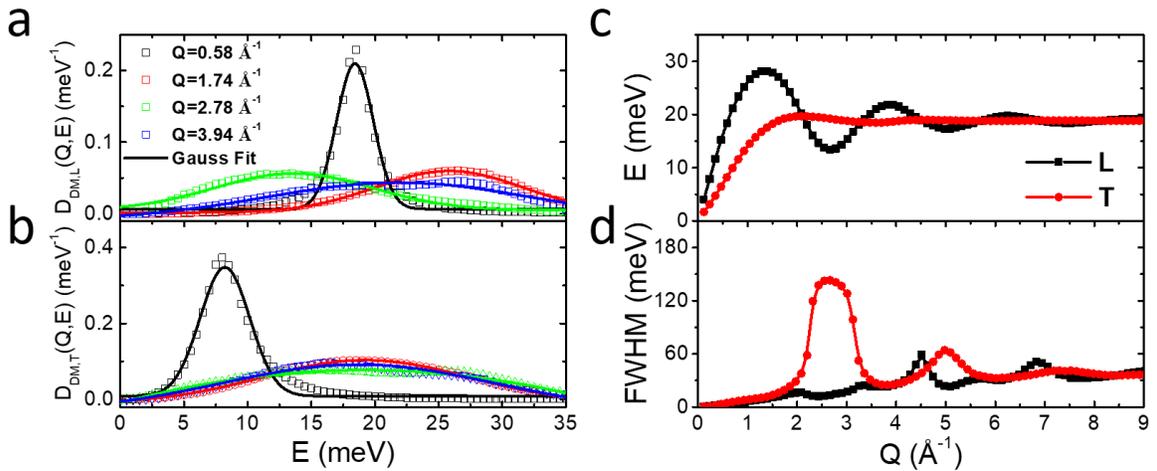

FIG. SI-9. The Gaussian fitting of longitudinal (a) and transverse (b) phonons at fixed $Q$ values, respectively. Four curves were presented here. (c) and (d) show the peak position and FWHM obtained from fitting for longitudinal and transverse phonons, respectively.

In the preceding work by Arai *et al.* [15], a Gaussian function was used to fit the peak in the *Q*-dependent GDOS in studying the phonon dispersion in a $Ni_{67}Zr_{33}$ MG. In this work, a Gaussian function and a Lorentzian function were adopted to fit the transverse and longitudinal VDOS to determine the peak positions and widths. The fitting results with a Gaussian function are demonstrated in FIG. SI-9 for MD spectra at selected *Q* values. The fitting results with a Lorentzian function are shown in FIG. SI-10.

Previous studies have demonstrated that the discussion of phonon linewidth by *Q*-dependent DOS is accurate at small *Q* values [12,16,17]. However, care should be taken for discussions of phonon peak widths at large *Q* values. With this caution in mind, the fitted phonon peak width can nevertheless provide a useful means to characterize the difference between longitudinal and transverse branches, for the following reasons: (1) For disordered materials, the structures are disordered and isotropic, so that only the magnitudes of the wave vector need to be considered. Therefore, there may not exist an ensemble of phonon modes with direction dependent distribution in *Q*-dependent DOS; (2) In the frequency regime of *E* <32 meV focused in this work, most modes are extended and there are no localized modes. As shown in FIG.3b, the localized modes are located in the high-frequency regime, *i.e.*, *E* >32 meV. Therefore, there is no ensemble of localized modes with non-localized ones in the frequency regime where the longitudinal and transverse branches were discussed.

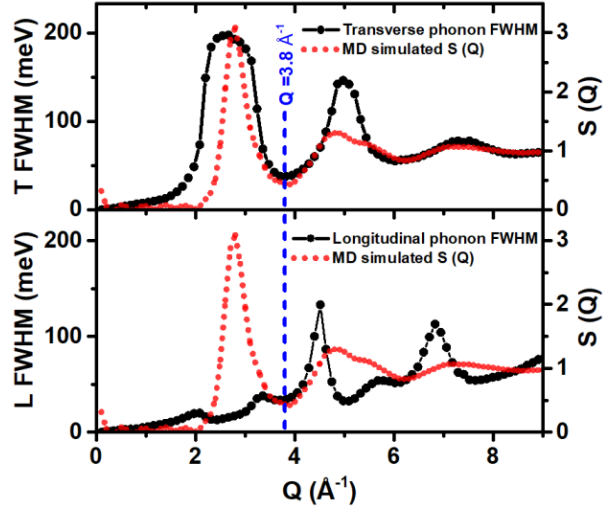

FIG. SI-10. Peak widths of the longitudinal and transverse phonon spectra vs. *S*(*Q*) obtained by MD simulations based on the DM method. The widths were obtained by fitting the MD spectra with a Lorentzian function. The one-to-one correlation can also be seen between the apparent peak width and *S*(*Q*) for the transverse phonon mode. These results are qualitatively similar to the Gaussian function fitting results (see FIG. 4a in manuscript). Again, there is no such correlation for the longitudinal phonon mode.

## S11. *Q*-dependent VDOS in the 3-dimensional Lennard-Jones (3DLJ) Glass

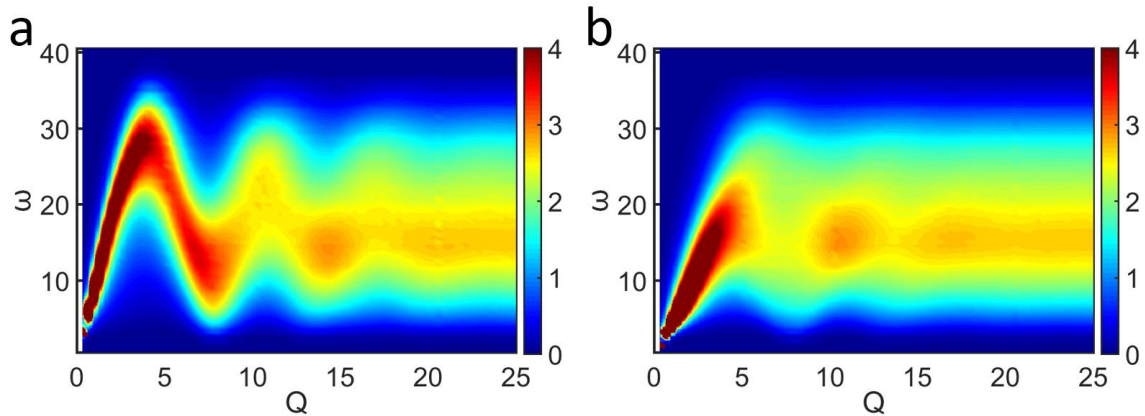

FIG. SI-11. *Q*-dependent VDOS of (a) longitudinal and (b) transverse phonons in a 3DLJ glass calculated by DM. The phonon spectra in the 3DLJ glass are consistent with those in $Zr_{46}Cu_{46}Al_8$ MG (see FIG. 2 e&f in manuscript).